\documentclass[aps,pra,twocolumn,notitlepage,superscriptaddress,showpacs,nofootinbib]{revtex4-1}

\usepackage{latexsym,amssymb,bm,bbold,amsfonts,amstext,graphicx,bbm,bm,relsize,dsfont,times}
\usepackage[dvipsnames]{xcolor}
\usepackage{enumerate}

\usepackage{amsmath}
\usepackage{amssymb}
\usepackage{amsfonts}
\usepackage{enumerate}
\usepackage{mathrsfs}
\usepackage{graphicx}
\usepackage{epstopdf}
\usepackage{enumerate}
\usepackage{changepage}

\newcommand{\tr}{\text{Tr}}

\usepackage[
colorlinks,
linkcolor = blue,
citecolor = blue,
urlcolor = blue]{hyperref}
\def \qed {\hfill \vrule height7pt width 7pt depth 0pt}

\newtheorem{theorem}{Theorem}
\newtheorem{definition}{Definition}
\newtheorem{lemma}{Lemma}

\begin{document}
	
	\title[]{Probabilistic and Approximate Masking of Quantum Information}
	
	\date{\today}
	
	\author{Mao-Sheng Li}
	\affiliation{Department of Physics,
			Southern University of Science and Technology, Shenzhen 518055, China}
		\affiliation{Department of Physics, University of Science and Technology of China, Hefei 230026, China}
	\author{Kavan Modi}
	\email{kavan.modi@monash.edu}
	\affiliation{School of Physics \& Astronomy, Monash University, Victoria 3800, Australia}
	\affiliation{Institute for Quantum Science and Engineering, and Department of Physics,	Southern University of Science and Technology, Shenzhen 518055, China}
	
\begin{abstract}
The no-masking theorem states that it is impossible to encode an arbitrary quantum state into the correlations between two subsystems so that no original information about is accessible in the marginal state of either subsystem. In this paper, we generalize this theorem to allow for failure of the protocol. We then bound the performance of a masking protocol when we are allowed a (probabilistic) approximate protocol.
\end{abstract}
 	\maketitle
	
\section{Introduction}

A quantum cloning machine, that can make perfect copies of an arbitrary quantum state, is forbidden by the structure of quantum mechanics~\cite{Wootter82, Dieks82}. This simple, yet profound, phenomenon plays a central role in many quantum information tasks, such as quantum teleportation~\cite{Bennett93, Bouwmeester97} and quantum key distribution~\cite{Bennett92, Gisin02}. This pioneering result has paved the way for a number of similar no-go theorems, such as the no-broadcasting theorem~\cite{Barnum96, Piani08, Kalev08}, the impossibility of a bit commitment protocol~\cite{Lo97, Mayers97}, the no-deleting theorem~\cite{Pati00}, and the no-hiding theorem~\cite{Pati07, Pati11}. On the other hand, the structure of quantum mechanics allows for several potential applications, such quantum key distribution~\cite{bb84, Ekert91} and secret sharing~\cite{Cleve99, Hillery99, DiVincenzo01}. In essence, all these phenomena are consequences of the superposition principle, along with the unitary evolution of quantum mechanics. Importantly, each theorem and task here sheds light on the boundary between the classical and the quantum worlds, which is of fundamental of interest in physics.

An important classical protocol that enables secrete communication is the one-time pad: here a classical message is combined with a random key (of the same size) so that neither the message nor the key contains the original information. However, the information if perfectly preserved in the correlations between the message and the key~\cite{Shannon49}. It is natural to consider the quantum analogue of the classical one-time pad. That is, can quantum information be hidden from	both subsystems and reside entirely in the correlation? Once again, the answer turns out to be no. Recently, Ref.~\cite{Kavan18} showed that this is impossible and named such phenomenon as the \textit{no-masking theorem}. As a by-product, this result implicates a \textit{quantum} version of the well-known impossibility of bit commitment protocols~\cite{Lo97, Mayers97}, i.e., the impossibility of a protocol that allows the commitment of qubits. However, it turns out that, while universal masking is impossible, the set of states that can be masked is non-trivial~\cite{Kavan18, Li18, Liang19, Ding19}. Moreover, when one allows for some trusted randomness in the protocol, the set of maskable states grows~\cite{Lie19_03, Lie19_08}, even to a point where universal masking becomes possible. These non-trivial features of the no-masking theorem has generated significant interest in the recent years~\cite{Yadin19, He19, Hotta19, Ghosh19, Guha19, Hotta20}.

When dealing with information-theoretic protocols, it is important to account for their robustness. That is, while making an exact copy of a quantum state is forbidden, is it also forbidden to make an approximate copy? This question was examined by several authors~\cite{Hillery96, Gisin97}, who found that it is possible to make imperfect copies of a quantum state. Similarly, others examined if probabilistic cloning of quantum states is possible~\cite{Duan98}. Then one natural question is to ask whether approximate or probabilistic quantum masking machines are allowed? If so, then some conditionally secure quantum qubit commitment could also be possible. In this paper, we prove that probabilistic universal masking is impossible and provide bounds on the achievable fidelity for an approximate universal masking machine. Finally, we show that this bound remains intact if we make the approximate masker also probabilistic. We begin by defining the masking protocol.
	
%%%%%%%%%%%%%%%%%%%%%%%%%%%%%%%%%%%%%%%%%%%%%
%%%%%%%%%%%%%%%%%%%%%%%%%%%%%%%%%%%%%%%%%%%%%
\section{Preliminaries}

The masking protocols involves two parties, $A$ and $B$. Let $\mathcal{H}_{A}$ $\mathcal{H}_{B}$ be Hilbert spaces of dimension $r\geq 2$ and $s\geq 2$, respectively. We denote the set of all pure states in $\mathcal{H}_{X}$ as $\mathbb{P} (\mathcal{H}_{X})$, while $\mathbb{D} (\mathcal{H}_{X})$ denotes all the density matrices of the system $X$, and the set $\mathcal{A} = \{|a_k\rangle_{A} \in \mathcal{H}_{A}\}$ is the set of maskable states.
	
\begin{definition} An operation $\mathcal{S}$ is said to mask quantum information contained in states $\mathcal{A}$ by mapping them to states $\{|\Psi_k\rangle_{AB}\in \mathcal{H}_{A} \bigotimes \mathcal{H}_{B} \}$ such that all the marginal states of $|\Psi_k\rangle_{AB}$ are identical, \emph{ i.e.},
\begin{gather}\label{eq:maskcond}
\rho_A \!=\! \tr_{B}(|\Psi_k\rangle_{\!AB} \langle \Psi_k|) \text{\emph{ and} } \rho_B \!=\! \tr_{A}(|\Psi_k\rangle_{\!AB} \langle \Psi_k|),
\end{gather}
\end{definition}
for all $k$. The masker can be modelled by a unitary operator $U_\mathcal{S}$ on $AB$, where $B$ is an ancillary system in a fixed state $\left|b \right>$:
\begin{gather}\label{eq:mask}
\mathcal{S}:\ \ U_\mathcal{S}|a_k\rangle_{A}|b\rangle_B =|\Psi_k\rangle_{AB}.
\end{gather}
One can think of the masker to be a triple $\mathcal{M}=( U_\mathcal{S}, \rho_A, \rho_B)$. An interesting goal is to determined the maximal set $S$ of states that can be masked by a given masker.
\begin{theorem}\label{nomasking}
\text{\emph{\cite{Kavan18}}} An arbitrary quantum state in $\mathbb{P}(\mathcal{H}_{A})$ cannot be masked.
\end{theorem}
By linearity the proof also applies for masking states in $\mathbb{D}(\mathcal{H}_A)$. The above theorem assumes that the initial state of $B$ is pure. This is because, as stated in Ref.~\cite{Kavan18}, for more than two parties masking is allowed~\cite{Hillery99, Cleve99}. Indeed Refs.~\cite{Lie19_03, Lie19_08} have shown that if this initial state is allowed to be mixed, i.e., there exists another system $C$ that purifies $B$, then universal masking is possible.
The no-masking theorem states there is no $\mathcal{M}$ that can mask all states in $\mathcal{H}_{A}$. Similarly to the no-cloning theorem, this impossibility also stems from the structure of quantum mechanics. However, for the no-cloning theorem if we do not demand that the cloning protocol works with certainty or precision it is possible to overcome the theorem, to some extent. Below, we extend the no-masking theorem to show the impossibility of probabilistic masking and bound the performance of an approximate masker.

%%%%%%%%%%%%%%%%%%%%%%%%%%%%%%%%%%%%%%%%%%%%%
%%%%%%%%%%%%%%%%%%%%%%%%%%%%%%%%%%%%%%%%%%%%%
%%%%%%%%%%%%%%%%%%%%%%%%%%%%%%%%%%%%%%%%%%%%%
\section{Impossibility of probabilistic universal masking} The first condition we relax is on the probability with which masking process succeeds. We can do this by replacing the unitary transformation $U_\mathcal{S}$ by a linear transformation $L_{\mathcal{S}}$.
	
\begin{definition}
An operation $\mathcal{S}_p$ is said to probabilistically mask quantum information contained in states of set $\mathcal{A}$ when $U_{\mathcal{S}}$ in Eq.~\eqref{eq:mask} is replaced by a a completely-positive and trace-decreasing linear transformation $L_{\mathcal{S}_p}$ that maps $\{|a_k\rangle_{A}|b\rangle_B\in\mathcal{H}_{A}\otimes\mathcal{H}_{B}\}$ to $\{p_k |\Psi_k\rangle_{AB}\}$ where $|\Psi_k \rangle_{AB} \in \mathbb{P} (\mathcal{H}_{A}\otimes \mathcal{H}_{B})$ such that all of the marginal states of $|\Psi_k\rangle_{AB}$ satisfy Eq.~\eqref{eq:maskcond}.
\end{definition}
	
Here, $L_{\mathcal{S}_p}$ is trace decreasing since the image occurs with a probability less than unity. In other words, this masking process is probabilistic, i.e., the process fails with probability $1-p_k$ for a given input $|a_k \rangle_A$. We also need to require that $L_{\mathcal{S}_p}$ is invertible, i.e., there exists mapping from $|\Psi_k\rangle_{AB} \to |a_k\rangle_A|b\rangle_B$. Clearly, such a machine will be less powerful than one that is guaranteed to achieve masking. We now show that $L_{\mathcal{S}_p}$ also cannot mask all states of $\mathbb{P} (\mathcal{H}_{A})$.
	
\begin{theorem}
\label{probabilistic_masking}
A probabilistic masker, that can mask all the states in $\mathbb{P}(\mathcal{H}_{A})$, is impossible.
\end{theorem}
	
	\noindent \emph{Proof.}
	We prove this claim by contradiction. Suppose $(L_{\mathcal{S}_p}, \rho_A,\rho_B)$ is such a process. Without loss of generality, we can assume $\rho_A=\sum_{j=1}^n\lambda_j|j\rangle\langle j|,$ with $\lambda_j>0$. Set
	\begin{gather}
	L_{\mathcal{S}_p}|1\rangle|b\rangle=p_1|\Psi_1\rangle,\ L_{\mathcal{S}_p}|2\rangle|b\rangle=p_2|\Psi_2\rangle.
	\end{gather}
	By assumption, $|\Psi_1\rangle, |\Psi_2\rangle$ are linearly independent and can be seen as two different purifications of $\rho_A$. Therefore, they can be written as the following form
	\begin{gather}\label{state}
	\begin{array}{c}
	|\Psi_1\rangle=\displaystyle\sum_{j=1}^{n} \sqrt{\lambda_j}|j\rangle |\mu_j\rangle, \ \
	|\Psi_2\rangle=\displaystyle\sum_{j=1}^{n} \sqrt{\lambda_j}|j\rangle |\nu_j\rangle,
	\end{array}
	\end{gather}
	where $\{|\mu_j\rangle\}_{j=1}^n$ and $\{|\nu_j\rangle\}_{j=1}^n$ are both orthonormal sets.
	
	Now for any pair $u_{ i }, v_{i}\in\mathbb{C}$, such that $|u_{i}|^2 + |v_{i}|^2=1$,
	\begin{gather}
	L_{\mathcal{S}_p}((u_{i}|1\rangle + v_{i} |2\rangle)|b\rangle) = u_{i} p_1|\Psi_1\rangle + v_{i} p_2|\Psi_2\rangle
	\end{gather}
	we get a nonzero vector as $L_{\mathcal{S}_p}$ is invertible. Let $N_i$ be the normalization for this vector. By assumption, the resulting states are purifications of $\rho_A$ and thus can be expressed as
	\begin{gather}\label{eq:phi}
	|\Phi_i\rangle:=
	\displaystyle\sum_{j=1}^n \sqrt{\lambda_j}|j\rangle \frac{u_i p_1|\mu_j\rangle+v_i p_2 |\nu_j\rangle} {N_i}.
	\end{gather}
	This means that the set of states $\{\frac{u_{i}p_1|\mu_j\rangle + v_{i} p_2 |\nu_j \rangle}{N_i}\}$ are orthonormal and for all $1\leq j\neq k\leq n$, we have $p_1 p_2 u_{i} \overline{v}_{i} \langle \nu_k|\mu_j\rangle+p_1 p_2 \overline{u}_{i} v_{i} \langle \mu_k|\nu_j\rangle=0$, here $\overline{u}_{i}$ means the complex conjugation of $u_{i}$.
	
	Particularly, choosing two sets $(u_1,v_1)$ and $(u_2,v_2)$ such that $u_1\overline{v_1} \overline{u}_2 {v_2} $ is not in $\mathbb{R}$,
	then we have
	\begin{gather}\label{Proba_mainequation}
	\begin{array}{l}
	u_1\overline{v_1}\langle \nu_k|\mu_j\rangle+ \overline{u_1} {v_1}\langle \mu_k|\nu_j\rangle=0,\\
	u_2\overline{v_2} \langle \nu_k| \mu_j\rangle + \overline{u_2} {v_2}\langle \mu_k| \nu_j\rangle=0.
	\end{array}
	\end{gather}
	The condition `$u_1\overline{v_1} \overline{u_2} {v_2} $ is not in $\mathbb{R}$' implies that the determinant of the above coefficients is nonzero. Therefore, we can deduce that $\langle \nu_k|\mu_j\rangle=0$ for all $j\neq k$. The states corresponding to the images of $L_{\mathcal{S}}$ can be also viewed as purifications of $\rho_B$ and by Eqs.~\eqref{state}, we have
	\begin{gather}\label{eq:rhoB}
	\sum_{j=1}^n\lambda_j|\mu_j\rangle\langle \mu_j| =\rho_B=\sum_{k=1}^n\lambda_k|\nu_k\rangle\langle \nu_k|.
	\end{gather}
	Calculating $\tr(\rho_B |\mu_l\rangle\langle \mu_l|)$ by substituting $\rho_B$ in two ways, we have $\lambda_l=\lambda_l|\langle \mu_l|\nu_l\rangle|^2$, hence $|\langle \mu_l|\nu_l\rangle|^2=1$ for $\lambda_l\neq 0$ and $|\nu_l\rangle= e^{i\theta_l}|\mu_l\rangle$ for some $\theta_l\in[0,2\pi]$ with $l=1,2,...,n$.

This turns the coefficient in the numerator of Eq.~\eqref{eq:phi} into $p_1u_i+p_2v_i e^{i\theta_j}$. Using the fact that the states in Eq.~\eqref{eq:phi} are also purification of $\rho_B$, we have equalities: $|p_1u_i+p_2v_i e^{i\theta_j}|^2 =|p_1u_i+p_2v_i e^{i\theta_k}|^2 $ for all $j, k\in \{1,2,...,n\}$. To see this, we trace out system $A$ of $|\Phi_i \rangle$, which gives us
\begin{gather}
\rho_B = \frac{1}{N_i^2}\sum_{j=1}^n \lambda_j |p_1u_i+p_2v_i e^{i\theta_j} |^2 | \mu_j\rangle\langle \mu_j|.
\end{gather}
Comparing this with the $\rho_B$ in Eq.~\eqref{eq:rhoB}, we must have $|p_1u_i+p_2v_i e^{i\theta_j} |^2 = N_i^2$ for all $j\in \{1,2,...,n\}$. This is equivalent to
\begin{gather}\label{prob_equation2}
\begin{array}{l} u_1\overline{v_1}(e^{-i\theta_j}-e^{-i\theta_k})+\overline{u_1} {v_1}(e^{i\theta_j}-e^{i\theta_k})=0，\\
u_2\overline{v_2}(e^{-i\theta_j}-e^{-i\theta_k})+\overline{u_2} {v_2}(e^{i\theta_j}-e^{i\theta_k})=0.
\end{array}
\end{gather}
The nonzero of determinant
$\left| \begin{array}{cc}	u_1\overline{v_1}&\overline{u_1} {v_1}\\	u_2\overline{v_2}&\overline{u_2} {v_2} \end{array}\right|$
gives us $e^{-i\theta_j} -e^{-i\theta_k} = e^{i\theta_j}-e^{i\theta_k}=0$ for all $1\leq j,k\leq n$, which simply means $|\Psi_2\rangle = e^{i\theta_1} |\Psi_1\rangle$. This is in contradiction with the assumption of linear independence of $|\Psi_1\rangle$ and $|\Psi_2\rangle$. \qed\\

\noindent{\bf{Remark:}} From the above proof, with the assumption that with $|\Psi_1\rangle$ and $|\Psi_2\rangle$ are linearly independent and $u_1\overline{v_1} \overline{u_2} {v_2} $ is not in $\mathbb{R}$, we can indeed obtain that four states $\{|\Psi_1\rangle, |\Psi_2\rangle, |\Phi_1\rangle, |\Phi_2\rangle\}$ cannot simultaneously purify $\rho_A, \rho_B$.

One can note that the probabilistic protocol here is different from the probabilistic cloning machine introduced in Ref.~\cite{Duan98}. There, the machine makes a measurement $M$ after a unitary transformation $U$ and the set of states there are restricted to be of finite. A probabilistic masker akin to Ref.~\cite{Duan98} is derived in Ref.~\cite{Lib19}. Our result goes in the converse direction and shows the impossibility of a universal probabilistic masker. Thus, giving up certainty for the success of the protocol does not help; perhaps giving up precision may help. We now ask can we have a masking process that can approximately mask all states in $\mathbb{P}(\mathcal{H}_A)$?

%%%%%%%%%%%%%%%%%%%%%%%%%%%%%%%%%%%%%%%%%%%%%
%%%%%%%%%%%%%%%%%%%%%%%%%%%%%%%%%%%%%%%%%%%%%
%%%%%%%%%%%%%%%%%%%%%%%%%%%%%%%%%%%%%%%%%%%%%
\section{Bounds on $\epsilon$-approximate universal masking} Physically, two states $\rho_A$ and $\rho'_A$ are said to be close to each other if we find it difficult to distinguish between them. Consider a masker maps a set of input states to states whose marginal states are close to a fiducial state. In this case, the information of the original states can only be obtained under some threshold hence approximate masking would be achieved. To formally quantify the distibguishiblity between two states we employ the quantum fidelity, $F(P,Q):= \tr \sqrt{\sqrt{P}Q\sqrt{P}}$, where $P,Q$ are two density matrices~\cite{nils, Watrous}. Notice that $0\leq F(P, Q)\leq 1$ where the greater the fidelity the more difficult it is to distinguish this two matrices $P$ and $Q$. With this we define an $\epsilon$-approximate masker as the following.
	
	\begin{definition} An operation $\mathcal{S}_\epsilon$ is said to $\epsilon$-approximately mask quantum information contained in states of set $ \mathcal{A}$ by mapping them to states $\{|\Psi_k\rangle_{AB}\in \mathcal{H}_{A} \bigotimes \mathcal{H}_{B} \}$ such that all marginal states of $|\Psi_k\rangle_{AB}$ approximately satisfy Eq.~\eqref{eq:maskcond}, i.e., they are only $\epsilon$ distinguishable from each other:
		\begin{gather}\label{approximate_condition}
		F(\rho_{A|k},\rho_{A|k'})\geq 1-\epsilon, \ \ \ F(\rho_{B|k},\rho_{B|k'})\geq 1-\epsilon,
		\end{gather}
		for all $ k,k'$. Here $\rho_{A|k}$ and $\rho_{B|k}$ are the marginals of $|\Psi_k\rangle_{AB}$.
	\end{definition}
	
An $\epsilon$-approximate masker can be designed by replacing $U_\mathcal{S}$ in Eq.~\eqref{eq:mask} with another unitary $U_\mathcal{\mathcal{S}_\epsilon}$ and demanding conditions~\eqref{approximate_condition}. Quantum fidelity is closely related to the trace norm $\big\| \bullet \big\|_1$ by the  inequality (see Ref. \cite{Watrous}, p161):
  $$ F(P,Q)\leq \sqrt{1-\frac{1}{4}\|P-Q\|_1^2}.$$
 If  the approximate conditions of Ineq.~(11) holds, the above inequality implies that
	\begin{gather}
	\big\|\rho_{X|k}-\rho_{X|k'}\big\|_1\leq 2\sqrt{2}\epsilon \quad \mbox{with} \quad X\in\{A,B\}.
	\end{gather}
	We now bound how well $\epsilon$-approximate masking is possible.
	
	\begin{theorem}
		\label{apprimate_masking_theorem}
		An approximate universal masker $\mathcal{S}_\epsilon$, that can mask all states in $\mathbb{P} (\mathcal{H}_{A})$ with fidelity greater than $1 - \epsilon$, can only do so for $\epsilon \ge \frac{\sqrt{2}}{72}(-1+ {\scriptstyle\sqrt{1+\frac{36} {\min\{r,s\}}}})$, where $r=\dim\mathcal{H}_A,s=\dim \mathcal{H}_B$.
	\end{theorem}
	\noindent\emph{Proof:} Set $t:=\min\{r,s\}.$ Let $\{|j\rangle_A \big | 1\leq j\leq r\}$ and $\{|k\rangle_B \big | 1\leq k\leq s\}$ be orthonormal basis of $\mathcal{H}_{A}$ and $\mathcal{H}_{B}$ respectively. Suppose that $ U_{\mathcal{S}_\epsilon} |1\rangle_A|b\rangle_B=|\Psi\rangle, \ U_{\mathcal{S}_\epsilon} |2\rangle_A |b\rangle_B =|\Phi\rangle$. The matrices representation of $|\Psi\rangle$ and $|\Phi\rangle$ are denoted by two matrices $M=(m_{jk}), N=(n_{jk})\in \text{Mat}_{r\times s}(\mathbb{C})$ respectively, i.e., $|\Psi\rangle=\sum_{j=1}^r\sum_{k=1}^s m_{jk}|j\rangle_A|k\rangle_B, \ \ |\Phi\rangle=\sum_{j=1}^r\sum_{k=1}^s n_{jk}|j\rangle_A|k\rangle_B.$ Then the partial traces of $|\Psi\rangle\langle\Psi|$ and $|\Phi\rangle\langle\Phi|$ can be calculated as follows:
	\begin{gather}\label{approx_trace}
	\begin{array}{cc}
	\tr_A(|\Psi\rangle\langle\Psi|)=M^{\dagger}M, \ \ \tr_A(|\Phi\rangle\langle\Phi|)= N^{\dagger}N; \\[2mm]
	\tr_B(|\Psi\rangle\langle\Psi|)=MM^{\dagger}, \ \ \tr_B(|\Phi\rangle\langle\Phi|)= NN^{\dagger}.
	\end{array}
	\end{gather}
	The definition of $\epsilon$-approximate quantum masking yields
	\begin{gather}\label{appximate_ineq_MN}
	\begin{array}{cc}
	\big\|MM^\dagger-NN^\dagger\big\|_1\leq 2\sqrt{2}\epsilon, \ \ \ \big\|M^\dagger M-N^\dagger N\big\|_1\leq 2\sqrt{2}\epsilon.
	\end{array}
	\end{gather}
	Moreover, considering another general state $|\Omega\rangle=u|\Psi\rangle+v|\Phi\rangle \in \mathbb{P}(\mathcal{H}_{A}\otimes \mathcal{H}_{B})$ for all $|u|^2+|v|^2=1$, we may find that the matrix corresponding to $|\Omega\rangle$ is just $uM+vN.$ Therefore, $\tr_B( |\Omega\rangle \langle\Omega|) = (uM+vN) (uM+vN)^\dagger $. By the definition of $\epsilon$-approximate quantum masking, we should also have $\big\| \tr_B( |\Omega\rangle\langle\Omega|)- \tr_B( |\Phi\rangle\langle\Phi|) \big\|_1\leq 2\sqrt{2}\epsilon.$ Writing it into the matrix form, we deduce the following inequality
	\begin{gather}
	\big\| |u|^2( MM^\dagger-NN^\dagger)+ u\overline{v}MN^\dagger+v\overline{u}NM^\dagger\big\|_1\leq 2\sqrt{2}\epsilon.
	\end{gather}
	By triangle inequality, $\big\|u\overline{v} MN^\dagger + v\overline{u}NM^\dagger\big\|_1 $ is less than
	\begin{align}
	\begin{split}
	& |u|^2\big\|( MM^\dagger-NN^\dagger)\big\|_1\\
	& + \big\| \ |u|^2( MM^\dagger-NN^\dagger)+ u\overline{v}MN^\dagger+v\overline{u}NM^\dagger\big\|_1
	\end{split}
	\end{align}
	which is less than $2\sqrt{2}(1+|u|^2)\epsilon$. Therefore,
	\begin{gather}
	\big\|u\overline{v}MN^\dagger+v\overline{u}NM^\dagger\big\|_1\leq 2\sqrt{2}(1+|u|^2)\epsilon.
	\end{gather}
	Choosing $(u,v)$ as the following pairs $(u_1,v_1)=(\frac{1}{\sqrt{2}},\frac{1}{\sqrt{2}})$ and $(u_2,v_2)=(\frac{i}{\sqrt{2}}, \frac{1}{\sqrt{2}})$ the above inequality becomes $\big\| \frac{1}{2}MN^\dagger \pm \frac{1}{2}NM^\dagger\big\|_1\leq 3\sqrt{2}\epsilon$. Using the triangle inequality of trace norm, we obtain $ \big\| MN^\dagger\big\|_1\leq 6\sqrt{2}\epsilon, $ and $ \big\|NM^\dagger\big\|_1\leq 6\sqrt{2}\epsilon.$ Using the compatible condition of norm, we obtain an upper bound of the modulo of $ \tr (MN^\dagger NM^\dagger)$. That is, $\!|\tr(MN^\dagger NM^\dagger)| \!\leq\! \big\|MN^\dagger NM^\dagger\big\|_1 \!\leq\! \big\|MN^\dagger\big\|_1\! \big\|NM^\dagger\big\|_1$, and the last term is less than $72\epsilon^2$.
	
	On the other hand, $\tr(MN^\dagger NM^\dagger)=\tr(M^\dagger MN^\dagger N)$. We will give a lower bound of $|\tr(M^\dagger MN^\dagger N)|$. Set $L=:N^\dagger N-M^\dagger M \in \text{Mat}_{s\times s}(\mathbb{C})$. Substituting $N^\dagger N=M^\dagger M+L$ into the above trace, we obtain
	\begin{gather}
	\begin{array}{ccl}
	\!\!|\tr(M^\dagger MN^\dagger N)|\!\!&=&\!\!|\tr((M^\dagger M)^2)\!+\!\tr(M^\dagger ML)| \\
	&\geq& \!\!\frac{1}{t}(\tr(M^\dagger M))^2 \!-\! |\tr(M^\dagger ML)|.
	\end{array}
	\end{gather}
	 We claim that $|\tr(M^\dagger ML)|\leq 2\sqrt{2}\epsilon$, which we prove below. Using this claim and the fact that $\tr(M^\dagger M)=1$, we have
	\begin{gather}
	|\tr(M^\dagger MN^\dagger N)|\geq \frac{1}{t}-|\tr(M^\dagger ML)|\geq \frac{1}{t}-2\sqrt{2}\epsilon.
	\end{gather}
	Hence we obtain the inequality which we need to prove
	\begin{gather}
	72\epsilon^2\geq | \tr(MN^\dagger NM^\dagger)|
	\!=\! |\tr(M^\dagger MN^\dagger N)|\geq \frac{1}{t}-2\sqrt{2}\epsilon. \notag
	\end{gather}
	Using the quadratic formula we find the lower bound on $\epsilon$ given in the statement of the theorem.
	
	Now we focus on proving the claim stated above. First, the second inequality of~\eqref{appximate_ineq_MN} implies that
	$\big\|L\big\|_2\leq \big\|L\big\|_1\leq 2\sqrt{2}\epsilon.$
	There exists some unitary matrix $U\in\text{U}(s)$ such that $UM^\dagger MU^\dagger$ is diagonal and the diagonals are $x_1,x_2,...,x_s$.
	Let $(y_1,y_2,...,y_s)$ be the diagonal elements of $ULU^\dagger.$ As $\big\|ULU^\dagger\big\|_2=\big\|L\big\|_2,$ we have
	\begin{gather*}
	\begin{array}{ccl}
	|\tr(M^\dagger ML)|&=&|\tr(UM^\dagger MU^\dagger UL U^\dagger)|\leq\sum_{j=1}^sx_j|y_j|\\
	&\leq &\sum_{j=1}^sx_j\big\|L\big\|_2=\big\|L\big\|_2\leq 2\sqrt{2}\epsilon. \qed
	\end{array}
	\end{gather*}
	
\noindent \textbf{Remark:} In the case $\epsilon=0$ for the above theorem, we can also arrive at the conclusion of the no-masking theorem.

Bu$\check{z}$ek and Hillery showed that a universal approximate cloning machine can clone of an arbitrary unknown qubit state with the surprisingly high fidelity of $\sqrt{5/6}\approx 0.913$~\cite{Hillery96}. Our bound here places restrictions on the performance of a universal masker in terms of the fidelity between the marginals of different masked states. It is interesting to compare our bound with that of Bu$\check{z}$ek and Hillery for the case of a qubit. In this case, the fidelity of the local states cannot greater than $0.6411$, which quantifies how much of the input information is revealed by the marginals.
	
%%%%%%%%%%%%%%%%%%%%%%%%%%%%%%%%%%%%%%%%%%%%%
%%%%%%%%%%%%%%%%%%%%%%%%%%%%%%%%%%%%%%%%%%%%%
%%%%%%%%%%%%%%%%%%%%%%%%%%%%%%%%%%%%%%%%%%%%%
\textbf{Probabilistic $\epsilon$-approximate masking.---} Before we make our conclusions, we look at the case where approximate masking is also allowed to be probabilistic. The next theorem is stronger statement contains the last theorem as a limiting case. Here, we bound the fidelity of an approximate masker if we allow for a probability of failure.

\begin{definition} An invertible operation $L_{\mathcal{S}_\epsilon}$ is said to probabilistic $\epsilon$-approximate masking quantum information contained in states $\mathcal{A} =\{|a_k\rangle_{A} \in\mathcal{H}_{A}\}$ by mapping them to $\{ p_k |\Psi_k \rangle_{AB}\in \mathcal{H}_{A}\bigotimes \mathcal{H}_{B} \}$ such that the marginal states of $|\Psi_k\rangle_{AB}$ cannot be distinguished with each other in the sense of $\epsilon$, i.e., they satisfy the approximate bounds in~\eqref{approximate_condition}.
\end{definition}

\begin{theorem}
\label{apprimate_probabilistic_masking_theorem}
An $\epsilon$-approximate-probabilistic masker $L_{\mathcal{S}_\epsilon}$ that can mask all states in $\mathbb{P}(\mathcal{H}_{A})$ with fidelity greater than $1-\epsilon$, can only do so for $ \epsilon \ge \frac{\sqrt{2}}{72}(-1+ {\scriptstyle\sqrt{1+\frac{36} {\min\{r,s\}}}})$.
\end{theorem}

\noindent\emph{Proof:} Since $L_{\mathcal{S}_\epsilon}$ is invertible, $\dim_\mathbb{C} (L_{\mathcal{S}_\epsilon} (\mathcal{H}_{A}\otimes |b\rangle )) = \dim_\mathbb{C} (\mathcal{H}_{A})=r\geq 2$. Hence there exist $|a_1\rangle, |a_2\rangle\in \mathbb{P} (\mathcal{H}_{A})$ such that $L_{\mathcal{S}_\epsilon}|a_j\rangle_A|b\rangle_B={p_j}|\Psi_j\rangle, j=1,2$ with $p_1,p_2>0$ and $\langle \Psi_1|\Psi_2\rangle=0$. The matrix representation of $|\Psi_j\rangle$ is denoted by the matrix $M_j=(m^{(j)}_{kl})\in \text{Mat}_{r\times s}(\mathbb{C})$. That is, $|\Psi_j\rangle=\sum_{k=1}^r\sum_{l=1}^s m^{(j)}_{kl}|k\rangle_A|l\rangle_B.$ Then we have the following partial traces
\begin{gather}\label{aprox_proba_trace}
\begin{array}{cc}
\tr_A(|\Psi_j\rangle\langle\Psi_j|)\!=\! M_j^{\dagger}M_j, \ \ \tr_B(|\Psi_j\rangle\langle\Psi_j|)\!=\! M_jM_j^{\dagger}.
\end{array}
\end{gather}
The definition of $\epsilon$-approximate probabilistic quantum masking yields
\begin{gather}\label{aprox_proba_inequa}
\begin{array}{cc}
\big\|M_1M_1^\dagger-M_2M_2^\dagger\big\|_1\leq 2\sqrt{2}\epsilon.
\end{array}
\end{gather}
	
	\begin{figure}[h]
\includegraphics[width=0.5\textwidth,height=0.5\textwidth]{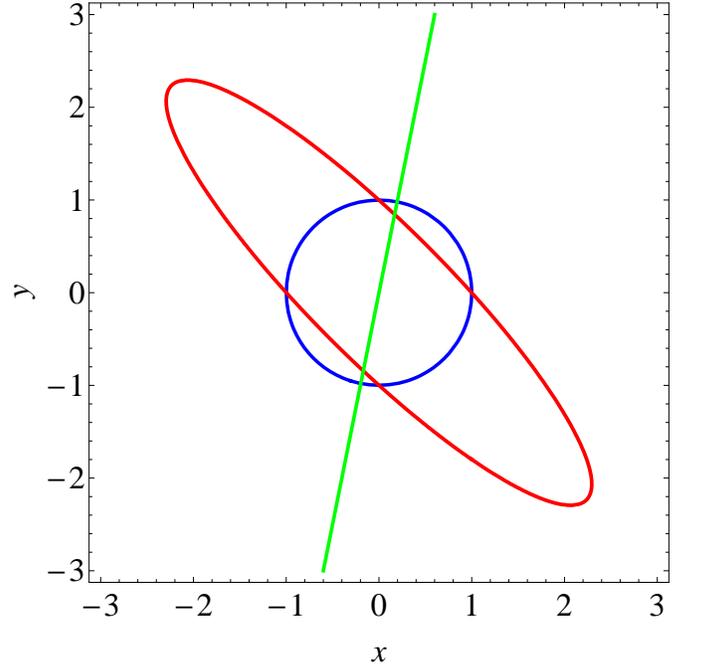}
\caption{Here are some examples for the curve $x^2+2r\cos\theta\ x y+y^2=1$. The blue circle corresponds to the parameters $r=0.9, \ \theta=\pi/2.$ The red ellipse corresponds to the parameters $r=0.9, \ \theta=0.$ And the line is $y=5x$.} \label{example_elliptic}
\end{figure}
We define $\mathcal{S}(\theta):=\{(x,y)\in\mathbb{R}^2 \ \big |\ x|a_1\rangle+y e^{i\theta} |a_2\rangle\in \mathbb{P}(\mathcal{H}_{A})\} $ for all $\theta\in[0,2\pi]$. For all $(x,y)\in\mathcal{S}(\theta)$, we have
\begin{gather*}
L_{\mathcal{S}_\epsilon}((x|a_1\rangle \!+\! y e^{i\theta}|a_2\rangle)|b\rangle_B) \!=\! \sqrt{(xp_1)^2 \!+\! (yp_2)^{2}}|\Omega_{x,y,\theta}\rangle,
\end{gather*}
$|\Omega_{x,y,\theta}\rangle=(xp_1|\Psi_1\rangle+yp_2e^{i\theta}|\Psi_2\rangle)/\sqrt{(xp_1)^2+(yp_2)^{2}}$ is a normalized state, where we have use the orthonormality of $|\Psi_1 \rangle$ and $|\Psi_2\rangle$ which plays a significant role in the proof. Lemma~\ref{particular_states}, given below, states that for any $\theta\in [0,2\pi]$ we can find some $(x,y)\in \mathcal{S}(\theta)$ such that $xp_1=yp_2$. For such parameters, $|\Omega_{x,y,\theta}\rangle=(|\Psi_1\rangle +e^{i\theta}|\Psi_2\rangle)/\sqrt{2}$. The matrix representation of such state $|\Omega_{x,y,\theta}\rangle$ is $(M_1+e^{i\theta}M_2)/\sqrt{2}.$ The partial trace $\tr_B(|\Omega_{x,y,\theta}\rangle\langle\Omega_{x,y,\theta}|)$ can be computed as
\begin{gather}
{\frac{1}{2}}(M_1M_1^\dagger+M_2M_2^\dagger+e^{-i\theta}M_1M_2^\dagger+e^{i\theta}M_2M_1^\dagger).
\end{gather}
The definition of $\epsilon$-approximate probabilistic quantum masking yields	$\big\| \tr_B(|\Omega_{x,y,\theta} \rangle \langle \Omega_{x,y,\theta}|) - M_1M_1^\dagger \big\|_1 \leq 2\sqrt{2}\epsilon$. That is,
\begin{gather}
\big\| M_2M_2^\dagger-M_1M_1^\dagger+e^{-i\theta}M_1M_2^\dagger+e^{i\theta}M_2M_1^\dagger\big\|_1\leq 4\sqrt{2}\epsilon.
\end{gather}
Then by Ineq.~\eqref{aprox_proba_inequa} and the triangle inequality, one has $\big\| e^{-i\theta} M_1 M_2^\dagger+e^{i\theta}M_2M_1^\dagger\big\|_1\leq 6\sqrt{2}\epsilon.$ As this holds for all $\theta\in[0,2\pi]$, particularly, we have
\begin{gather}
\begin{array}{l}
\big\|M_1M_2^\dagger \pm M_2M_1^\dagger\big\|_1\leq 6\sqrt{2}\epsilon.
\end{array}
\end{gather}
Therefore, by using triangle inequality again, we deduce $\big\|M_1M_2^\dagger\big\|_1\leq 6\sqrt{2}\epsilon, \ \big\|M_2M_1^\dagger\big\|_1\leq 6\sqrt{2}\epsilon.$ Using the compatible condition of norm, we obtain an upper bound of the modulo of $\tr(M_1M_2^\dagger M_2M_1^\dagger)$. That is,
\begin{gather}\label{main_inequa}
\begin{array}{c}
\big\|M_1M_2^\dagger M_2M_1^\dagger\big\|_1\leq \big\|M_1M_2^\dagger\big\|_1\big\|M_2M_1^\dagger\big\|_1 \leq 72 \epsilon^2.
\end{array}
\end{gather}
Then we can take similar argument as the proof of Theorem~\ref{apprimate_masking_theorem} to complete the proof and therefore we omit it here. \qed \\
	
%%%%%%%%%%%%%%%%%%%%%%%%%%%%%%%%%%%%%%%%%%%%%
%%%%%%%%%%%%%%%%%%%%%%%%%%%%%%%%%%%%%%%%%%%%%
%%%%%%%%%%%%%%%%%%%%%%%%%%%%%%%%%%%%%%%%%%%%%
\section{Conclusions} In this paper, we have considered three generalisations of the no-masking theorem. First, by replacing the unitary operation with an invertible linear operation we obtain the impossibility of probabilistic universal masking. While giving up certainty does not help for masking, we find that giving up the precision does. We derive a necessary condition of the precision $\epsilon$ to make the approximate masking possible. At last, we combine the previous process together and consider the $\epsilon$-approximate-probabilistic masker. These results can be seen as the robustness of the no-masking theorem. On the other hand, our results and methods open up new research questions, namely, finding optimal approximate maskers and testing the tightness of our fidelity bounds. The implications of our results the subsequent open questions have important applications quantum secret sharing, data hiding, and other quantum information protocols that require storing of information in composite quantum systems. Some important examples include the studies of interacting prover problems~\cite{Fitzsimons}, quantum error correction methods, and
\textit{out of time order correlators}, which are important for the black hole physics and the black hole information paradox.

\appendix
	
\section{Appendix}

\begin{lemma}\label{particular_states}
Let $|\alpha_1\rangle, |\alpha_2\rangle\in \mathbb{P}{(\mathcal{H}_A)}$ and $0 \leq \langle \alpha_1|\alpha_2\rangle<1$. For any positive real numbers $p_1,p_2\in \mathbb{R}_+$ and $\theta\in[0,2\pi]$, there exists $x, y\in \mathbb{R}$ such that $x|\alpha_1\rangle+y e^{i\theta}|\alpha_2\rangle\in \mathbb{P}{(\mathcal{H}_A)}$ and $p_1x=p_2y$.
\end{lemma}
\noindent \emph{Proof.} Denote $r:=\langle \alpha_1|\alpha_2\rangle$. The vector $x|\alpha_1\rangle+y e^{i\theta}|\alpha_2\rangle\in \mathbb{P}{(\mathcal{H}_A)}$ if and only if
\begin{gather}\label{elliptic}
x^2+2r\cos\theta\ x y+y^2=1.
\end{gather}
Fixed $r$ and $\theta$, the above function defines an elliptic curve whose center is $(0,0)$. In fact, we can write it into the following normal form:
\begin{gather}
\frac{(x+r\cos \theta \ y)^2}{1}+\frac{y^2}{(1-r^2 \cos^2 \theta)^{-1}}=1.
\end{gather}
Obviously, the curve defined by Eq.~\eqref{elliptic} has incidents with the line defined by
\begin{gather}\label{linear}
p_1 x=p_2 y.
\end{gather}
Equivalently, there is some common solution of the Eqs.~\eqref{elliptic} and \eqref{linear} as shown in Fig.~\ref{example_elliptic}. This completes the proof. \qed

\begin{acknowledgements}
{\bf Acknowledgements.} We thank J. Fitzsimons for bringing our attention to this problem. M.-S. L. thanks Xiangjing Liu for helpful discussions. K.M. received travel support from the \textit{Australian Academy of Technology and Engineering} via the 2018 \textit{Australia China Young Scientists Exchange Program} and the 2019 \textit{Next Step Initiative}. M.-S. L. would like also to thank the National Natural Science Foundation of China (Grants
No. 11875160 and No. 11871295) for partial financial support.
\end{acknowledgements}

\end{document}